\documentclass[a4paper,12pt]{article}

\usepackage{graphicx}
\usepackage{harvard}
\citationmode{abbr}
\begin{document}

\begin{center}
\textbf{Intuitive calculation of the relativistic Rayleigh-Taylor instability linear growth rate}\\
Antoine Bret\\

ETSI Industriales, Universidad de Castilla-La Mancha,\\
13071 Ciudad Real, Spain\\
\end{center}

\bigskip

\begin{abstract}
The Rayleigh-Taylor instability is a key process in many fields of Physics ranging from astrophysics to inertial confinement fusion. It  is usually analyzed deriving the linearized fluid equations, but the physics behind the instability is not always clear. Recent works on this instability allow for an very intuitive understanding of the phenomenon and for a straightforward calculation of the linear growth rate. In this Letter, it is shown that the same reasoning allows for a direct derivation of the relativistic expression of the linear growth rate for an incompressible fluid.
\end{abstract}


\newpage

\section{Introduction}
``What I cannot create, I do not understand'' was once found written on Richard Feynman's blackboard. Albert Einstein stated that ``You do not really understand something unless you can explain it to your grandmother.'' The point made by these two great minds was that understanding something in physics means you come to the point when it seems obvious and you no longer need the equations to derive the result. Among the physical ideas that have been examined over the years, the stability/instability concept has been extremely fruitful, although its understanding in a given setting would not always win Feynman's or Einstein's approval. The stability of a ball inside a bowl is very intuitive, and its oscillation frequency when removed from its equilibrium position can be calculated just looking at a sketch of the system. The \emph{instability} of a pencil on its tip is equally obvious and frequently cited when introducing the concept of an unstable system. In plasma physics, Fried could analyze the filamentation instability of two counter-propagating particle beams from the very understanding of the physical mechanism at work \cite{Fried1959}. Yet, a similar intuitive derivation for the  well-known two-stream instability is still lacking.

 The Rayleigh-Taylor instability (RTI) plays a key role in many fields of physics, and its behavior in connection with inertial confinement  fusion (ICF) \cite{LPBRT1990,KawataLPB1993}, $z$-pinch physics \cite{DouglasLPB2001} or metallic hydrogen generation experiments \cite{PirizLPB2006,JJLPB2006} has been the topic of many recent works.  The RTI occurs when a heavy fluids is accelerated against a light fluid (see Figure \ref{fig:rt}). In ICF,  a spherical Deuterium-Tritium target is compressed by a Laser. The laser ablates the target, creating a low density ablating plasma outside the pellet. During the early phase of the compression, the interface between  the compressed target (the heavy fluid) and the low-density ablating plasma (the light fluid) accelerates \cite{PirizSanz}. An observer ``sitting'' on the interface would then feel a force pushing him from the heavy fluid to the light one, resulting in the RT unstable configuration pictured on the Figure. In astrophysics, the RTI is frequently invoked to explain the filamentary structure of the Crab nebula for example \cite{HesterRTI}. As the supernova remnant (the dense fluid) decelerates through the interstellar medium (the light fluid), the interface between both is again RT unstable as it experiences an acceleration from the heavy to the light medium.

 A Feynman/Einstein like heuristic approach to the RTI \cite{Rayleigh,Taylor1950} was recently provided by Piriz \emph{et. al.} \cite{PirizAjp} for an incompressible fluid. In the usual, normal modes approach, where the fluid equations for both fluids are linearized, the linear growth rate is derived but the basic mechanisms at works remain hidden behind the equations \cite{ChandraFluid}. In contrast, Piriz \emph{et. al.}  proposed a direct derivation of the linear growth rate from the very description of the physics involved, ``short-cutting'' much of the equations.

\section{Non-relativistic approach}
\begin{center}
\begin{figure}
\includegraphics[width=0.9\textwidth]{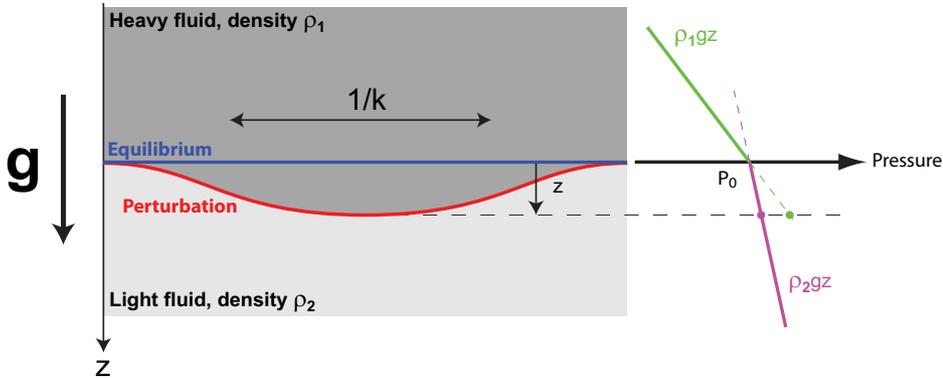}
\caption{The Rayleigh-Taylor instability. At equilibrium, the pressure on both side of the interface is $p_0$. The pressure variation when moving the interface shows the perturbation is unstable if $\rho_1>\rho_2$.}
\label{fig:rt}
\end{figure}
\end{center}

Suppose the interface represented on Figure \ref{fig:rt} is initially in equilibrium, both incompressible fluids exerting a pressure $p_0$ upon it. The system is accelerated downward with an intensity $g$ m/s$^2$. The interface is then displaced by $z$ along a distance $\sim 1/k$, where $k$ mimics here the wave-number introduced in the normal modes approach. The pressure of the upper-fluid at the interface increases by an amount $\rho_1 g z$, while the pressure of the lower fluid also increases, but by a quantity $\rho_2 g z$. The upper fluid now pushes the interface downward with the pressure $p_0+\rho_1 g z$, and the lower fluid pushes upward with the pressure $p_0+\rho_2 g z$. It is obvious that if $p_0+\rho_1 g z > p_0+\rho_2 g z$, i.e. $\rho_1>\rho_2$, the pressure balance amplifies the perturbation. Note that according to a similar analysis, moving the interface upward equally triggers an instability.

The calculation of the linear growth rate is straightforward from this stage. Let us consider a transverse direction, say $y$, to Fig. \ref{fig:rt} so as to account for the 3D nature of the system. The surface of the interface over a depth $D$ along the $y$ axis is $S\sim D/k$. The force acting upon it thus reads,
\begin{equation}\label{eq:F}
    F\sim (\rho_1-\rho_2)g z S = (\rho_1-\rho_2)g z \frac{D}{k}.
\end{equation}
As it moves, the interface also moves a layer of fluid on both sides. The volume of fluid displaced over the surface $S$ should be proportional to $S$ itself, and the height of the layer moved on both side is proportional to $1/k$ (which is why the normal mode calculation assumes the fluid thickness is much larger than $1/k$). We can thus write the expression of the mass involved in the displacement,
\begin{equation}\label{eq:M}
   M \propto \rho_1 S/k+\rho_2 S/k = \frac{\rho_1+\rho_2}{k^2}D.
\end{equation}
Since this amount of fluid is displaced by a distance $z$, we can write Newton's law $\mathbf{F}=M\mathbf{a}$ from Eqs.(\ref{eq:F},\ref{eq:M}) as,
\begin{equation}\label{eq:Newton}
   (\rho_1-\rho_2)g z \frac{D}{k} = \frac{\rho_1+\rho_2}{k^2}D\frac{d^2z}{dt^2}\Leftrightarrow \delta^2 z = \frac{d^2z}{dt^2},
\end{equation}
where,
\begin{equation}\label{eq:delta}
   \delta^2 = \frac{\rho_1-\rho_2}{\rho_1+\rho_2}g k,
\end{equation}
and a proportionality coefficient equal to unity has been assumed for the mass in Eq. (\ref{eq:M}). Equation (\ref{eq:Newton}) has exponentially growing solutions if $\delta^2>0$, and $\delta$ is exactly the linear growth rate of the RTI \cite{Rayleigh} where the Atwood number $A_t=(\rho_1-\rho_2)/(\rho_1+\rho_2)$ is immediately identified. It is clear now that the sum of the densities relates to the total amount of fluid involved in the motion, while the difference relates to the pressure shift generated by the displacement. Note that a mass factor different from unity in Eq. (\ref{eq:M}) would yield a slightly different denominator for the linear growth rate. Indeed, such intuitive calculations frequently yield the correct scalings with some pre-factor close to unity. In the present case, the result is exact.

\section{Relativistic version}
The relativistic version of the RTI is especially relevant to Supernovae and Gamma-Ray-Bursts physics \cite{WaxmanPiran,Levinsona}, where ultra-relativistic inhomogeneous flows are involved. Adapting the normal modes method to such settings, Allen \& Hugues \cite{AllenHugues} found the  relativistic counterpart of Eq. (\ref{eq:delta}),
\begin{equation}\label{eq:delta_R}
   \delta^2 = \frac{\rho_1-\rho_2}{8p_0/c^2+\rho_1+\rho_2}g k.
\end{equation}

Let us now analyze the problem from the intuitive standpoint explained above. Starting from Eq. (\ref{eq:delta}), where exactly shall we have to introduce relativistic expressions? The displacement itself is not relativistic. The interface corrugation is a still, initial condition. The only relativistic modification will have to do with the inertia of the fluid displaced. A volume of fluid $dV$ has the mass $\rho dV$ in the non-relativistic limit. If the particles it is made of have relativistic motion, for example a temperature $T$ such as $k_BT\sim mc^2$, the energy $p dV$ adds up to the mass within an amount $\propto p dV/c^2$. Relativistic fluid theory shows indeed that the correct factor is 3 so that the relativistic mass density is $\rho+3p/c^2$ \cite{landau2009}. The density term in Eq. (\ref{eq:M}) needs therefore to account for this extra inertia. While the interface has not been displaced, the pressure is the same on both side and the correction per unit of volume reads  $3p_0/c^2$ for both fluids. Updating the pressure here for the corrugated interface would introduce a second order term in $z$, which is neglected in the present linear regime. The relativistic counterpart of Eq. (\ref{eq:M}) is thus readily obtained replacing $\rho_1+\rho_2$ by $\rho_1+3p_0/c^2+\rho_2+3p_0/c^2$, and the new linear growth rate is,
\begin{equation}\label{eq:delta_R1}
   \delta^2 = \frac{\rho_1-\rho_2}{6p_0/c^2+\rho_1+\rho_2}g k.
\end{equation}
The  calculation starting from the linearized relativistic fluid equations \cite{AllenHugues} yields the same expression with a term $8p_0/c^2$  instead of $6p_0/c^2$.

\section{Discussion}
Equation (\ref{eq:delta_R1}) thus give, up to a numerical factor, the correct value of the RTI linear growth rate. We now discuss the discrepancy between the factors 6 and 8. To do so, we can start from the relativistic Euler equation in the absence of gravitational field \cite{LandauFluid},
\begin{equation}\label{eq:Euler}
(p+\varepsilon)u^k\frac{\partial u_i}{\partial x^k}=\frac{\partial p}{\partial x^i}-u_iu^k\frac{\partial p}{\partial x^k},
\end{equation}
where $p$ is the pressure, $e$ the energy density, $x^i=(ct,\mathbf{r})$ and $u^i=(\gamma,\gamma \mathbf{v}/c)$. As previously said, our problem is relativistic in the sense that the energy density can be so, not by virtue of some relativistic velocity of the fluid elements. Setting thus $\gamma=1$, one can check that the temporal component ($i=0$) of the equation above yields $\mathbf{v}\cdot \nabla p=0$. The spatial part ($i=1,2,3$) then gives
\begin{equation}
\frac{\mathbf{v}}{c^2}\frac{\partial p}{\partial t}+\nabla p
=
-\frac{p+\varepsilon}{c^2}\left(\frac{\partial}{\partial t}+\mathbf{v}\cdot\nabla\right) \mathbf{v}.
\end{equation}
Neglecting $\mathbf{v}\cdot\nabla \mathbf{v}$ as a second order quantity and adding the acceleration gives the premise of Allen \& Hugues' Eq. (3),
\begin{equation}\label{eq:OK}
\frac{\mathbf{v}}{c^2}\frac{\partial p}{\partial t}+\nabla p
=
-\frac{p+\varepsilon}{c^2}\frac{\partial\mathbf{v}}{\partial t}-g(p+e).
\end{equation}
Equations (\ref{eq:Euler},\ref{eq:OK}) show that that the correct relativistic inertia is not simply the energy density $\varepsilon$, but the energy density plus the pressure, $\varepsilon+p$. Setting then $\varepsilon = \rho c^2+3p$ gives $p+\varepsilon = \rho c^2+4p$ for both fluids, from which the factor 8 eventually arises.

\section{Acknowledgements}
The author acknowledges the financial support y Projects
ENE2009-09276 of the Spanish Ministerio de Educaci\'{o}n y Ciencia and PAI08-0182-3162 of the Consejer\'{i}a de Educaci\'{o}n y Ciencia de
la Junta de Comunidades de Castilla-La Mancha. Thanks are due to  Roberto Piriz  for fruitful discussions.

\bibliographystyle{agsm}
\bibliography{BibBret}

\end{document}